\documentclass[twocolumn,epjc3]{svjour3}
\smartqed  
\usepackage{preprintcover}  
\PreprintCoverPaperTitle{Search for a fermiophobic Higgs boson in the diphoton decay channel with the ATLAS detector}  
\PreprintIdNumber{CERN-PH-EP-2012-105}  
\PreprintCoverAbstract{A search for a fermiophobic Higgs boson using diphoton events produced in proton-proton collisions at a centre-of-mass energy of $\sqrt{s}=7$~TeV is performed using data corresponding to an integrated luminosity of 4.9~\ifb\ collected by the ATLAS experiment at the Large Hadron Collider. A specific benchmark model is considered where all the fermion couplings to the Higgs boson are set to zero and the bosonic couplings are kept at the Standard Model values (fermiophobic Higgs model). The largest excess with respect to the background-only hypothesis is found at 125.5~GeV, with a local significance of 2.9 standard deviations, \linebreak which reduces to 1.6 standard deviations when taking into account the look-elsewhere effect. The data exclude the fermiophobic Higgs model in the ranges 110.0 -- 118.0~GeV and 119.5 -- 121.0~GeV at 95\% confidence level. }  
\PreprintJournalName{Eur. Phys. J. C}  

\RequirePackage{fix-cm}
\RequirePackage{lineno}
\RequirePackage{graphicx}
\RequirePackage{dcolumn}
\RequirePackage{bm}
\RequirePackage[hyperindex,breaklinks]{hyperref}
\RequirePackage{placeins}
\RequirePackage{multirow}
\RequirePackage{subfig}
\RequirePackage{cite}

\newcommand{\hgg}{$H\rightarrow~\gamma\gamma$}
\newcommand{\infb}{fb$^{-1}$}
\newcommand{\lumi}{4.9~\infb}
\newcommand\T{\rule{0pt}{2.6ex}}
\newcommand\B{\rule[-1.2ex]{0pt}{0pt}}

\newcommand{\myfontsizetwo}{\fontsize{8}{10}\selectfont}

\newcommand{\mgg}{\ensuremath{m_{\gamma\gamma}}}
\newcommand{\ptt}{\ensuremath{p_{\mathrm{Tt}}}}
\newcommand{\ifb}{fb$^{-1}$}
\newcommand{\pt}{\ensuremath{p_{\mathrm{T}}}}
\newcommand{\Et}{\ensuremath{E_{\mathrm{T}}}}

\newcommand{\Zee}{\ensuremath{Z \rightarrow e^+e^-}}

\newcommand{\mH}{\ensuremath{m_{H}}}

\journalname{Eur. Phys. J. C}

\begin{document}

\title{Search for a fermiophobic Higgs boson in the diphoton decay channel with the ATLAS detector}

\author{The ATLAS Collaboration}

\date{\today}

\maketitle

\hyphenation{ATLAS}

\begin{abstract}
  A search for a fermiophobic Higgs boson using diphoton events produced in proton-proton collisions at a centre-of-mass energy of $\sqrt{s}=7$~TeV is performed using data corresponding to an integrated
   luminosity of 4.9~\ifb\ collected by the ATLAS experiment at the Large Hadron Collider. A specific benchmark model is considered where all the fermion couplings to the Higgs boson are set to zero and the bosonic couplings are kept at the Standard Model values (fermiophobic Higgs model). The largest excess with respect to the background-only hypothesis is found at 125.5~GeV, with a local significance of 2.9 standard deviations, \linebreak which reduces to 1.6 standard deviations when taking into account the look-elsewhere effect.
   The data exclude the fermiophobic Higgs model in the ranges 110.0 -- 118.0~GeV and 119.5 -- 121.0~GeV at 95\% confidence level.


\end{abstract}

Several extensions of the Standard Model (SM) have
been proposed in which the Higgs field couplings to some
or all fermion generations are substantially suppressed, for example two Higgs doublet models or Higgs triplet models~\cite{Bamert:1993ah,Barroso:1999bf,Gunion:1989ci,Akeroyd:2010eg}.
A fermiophobic benchmark model, in which the Higgs field couplings to all fermions are set to zero while the couplings to bosons are kept at their SM values, has been introduced to allow a generic investigation of these scenarios~\cite{lep:2001yc}.

In such a model, the production of the Higgs boson in hadron colliders and its decay properties are significantly altered compared to the SM.
Fermiophobic Higgs bosons can only be produced via vector boson fusion (VBF) or associated production with vector bosons (VH, $\mbox{V} = W,Z$). Because
Higgs boson decays to fermions are absent at tree level, the branching
fractions for decays to gauge bosons are enhanced. In addition,
the partial width of the decay to two photons is enhanced by the
suppression of the destructive interference between the $W$-boson
and top-quark loops.
The resulting cross section times branching ratio for fermiophobic Higgs boson production with decay to two photons is larger than that of the SM for Higgs boson masses (\mH) below 125~GeV. Table~\ref{tab:signal} lists, for several values of \mH, the fermiophobic Higgs boson cross section multiplied by the decay branching ratio into two photons. The ratio of this quantity with respect to that of the SM Higgs boson and the enhancement of the diphoton branching ratio are also shown. In
addition to the enhanced diphoton decay rates, the recoiling jets
or vector bosons in the VBF or VH production modes, respectively, imply a high transverse momentum for the Higgs boson that can be exploited as a discriminating variable in the analysis. However, for increasing \mH\ the diphoton decay rate falls rapidly, making the search less sensitive at higher masses in this decay channel.

\begin{table}[t!]
\caption{\label{tab:signal}Higgs boson production cross section multiplied by the branching ratio into two photons for the fermiophobic benchmark model ($\sigma_\mathrm{f}$), the ratio of this value to the SM value ($\sigma_\mathrm{f} / \sigma_\mathrm{SM}$) and the two photon branching ratio enhancement compared to the SM ($\cal{B}_\mathrm{f} / \cal{B}_\mathrm{SM}$) for various fermiophobic Higgs boson masses.
  The expected number of signal events after candidate selection are also shown for \lumi\ of data as well as the overall signal selection efficiencies.}
\begin{center}
{\myfontsizetwo
\begin{tabular}{l@{\phantom{....}}c@{\phantom{..}}c@{\phantom{..}}c@{\phantom{..}}c@{\phantom{..}}c@{\phantom{..}}c@{\phantom{..}}c@{\phantom{..}}c@{\phantom{..}}c}
\hline\hline
\T \B \mH\ [GeV]            & 110 & 115 & 120 &125 &130 &135 &140 &145 &150  \\
\hline
\T \B $\sigma_\mathrm{f}$ [fb]  & 163 & 90 & 53 & 32 & 21 & 13 & 8.9 & 5.9 & 3.9 \\
\T \B $\sigma_\mathrm{f} / \sigma_\mathrm{SM}$ & 3.7 & 2.1 & 1.2 & 0.8 & 0.6 & 0.4 & 0.3 & 0.3 & 0.2 \\
\T \B $\cal{B}_\mathrm{f} / \cal{B}_\mathrm{SM}$ & 30.2 & 17.0 & 10.3 & 6.7 & 4.7 & 3.5 & 2.8 & 2.3 & 2.0 \\
\T \B Signal events           & 255 & 149 & 91 & 58 & 38 & 25 & 17 & 12 & 7.9 \\
\T \B Efficiency [\%]         & 32 & 34 & 35 & 37 & 38 & 38 & 39 & 40 & 42 \\
\hline\hline
\end{tabular}}
\end{center}
\end{table}

Searches for a fermiophobic Higgs boson have been performed at the LEP and Tevatron colliders. The combination of results from the LEP experiments~\cite{lep:2001yc} excludes a fermiophobic Higgs boson at 95\% confidence level (CL) for masses below 109~GeV. When including both the $WW$ and $\gamma\gamma$ decay modes, the Tevatron experiments exclude a fermiophobic
Higgs boson with masses up to 119~GeV~\cite{Abazov:2011ix,A:2011tg}.

This letter describes a search for a fermiophobic Higgs boson using diphoton events produced in proton-proton collisions at a centre-of-mass energy of $\sqrt{s} = 7$~TeV using data corresponding to an integrated
luminosity of 4.9~\ifb\ collected by the ATLAS experiment. This analysis follows exactly that of the related search for a SM Higgs boson with the same dataset~\cite{Hgg_paper}, but the fermiophobic Higgs hypothesis is used to construct the signal model. The sensitivity
to the fermiophobic signal is larger than that for the SM Higgs due to the larger diphoton transverse momentum.

The ATLAS detector is described in detail in Ref.~\cite{ATLAS_detector}. The most relevant subsystems for this analysis are the
calorimeter, in particular the electromagnetic section, and the inner detector. The electromagnetic calorimeter is a lead--liquid-argon detector, finely  segmented in the lateral and longitudinal directions. It is composed of a barrel part covering the pseudorapidity
range $|\eta| < 1.475$ and two end-cap sections covering $1.375 < |\eta| < 3.2$.
The barrel ($|\eta| < 0.8$) and extended barrel ($0.8 < |\eta| < 1.7$) hadron calorimeter sections consist of steel and scintillating tiles, while the end-cap sections ($1.5 < |\eta| < 3.2$) are composed of copper and liquid argon.
The inner detector includes silicon-based pixel and micro-strip detectors in the range $|\eta| < 2.5$, and a transition radiation tracker with electron identification capability extending out to $|\eta| < 2.0$. It is surrounded by a superconducting solenoid that provides a 2~T axial magnetic field.

Data used in this analysis were recorded using a diphoton trigger with a 20~GeV transverse energy (\Et) threshold on each photon. This trigger is seeded by a first-level trigger, which requires two clusters in the electromagnetic calorimeter with $\Et > 14$~GeV or $\Et > 12$~GeV, depending on the data-taking period.
This trigger has a signal efficiency close to 99\% following the
final event selection.
After application of data-quality requirements the analysed data sample corresponds to a total integrated luminosity of $4.9\pm 0.2$~\infb \cite{lumiAll}.

The events are required to have at least one reconstructed vertex with a minimum of three associated tracks, where the transverse momentum of each track is required to be larger than 0.4~GeV. At least two photons within the fiducial region $|\eta| < 2.37$ (excluding the transition region between the barrel and the end-cap,  $1.37 < |\eta| < 1.52$) satisfying tight identification criteria based on electromagnetic shower shapes~\cite{ATLAS:photons} are required. The transverse momenta for the leading and sub-leading photons are required to be larger than 40~GeV and 25~GeV, respectively. The photon reconstruction and identification efficiency ranges typically from 65\% to 95\% for $E_{\rm{T}}$ in the range between 25~GeV and 80~GeV.
The transverse energy deposited around each photon within a cone of $\Delta R = \sqrt{\left( \Delta \eta \right)^2 + \left( \Delta \phi \right)^2} = 0.4$, excluding the deposits of the photon itself, is required to be less than 5~GeV. Corrections for the small estimated energy leakage outside the excluded region, the underlying event and effects of additional minimum bias interactions occuring in the same or neighbouring bunch crossings (in-time and out-of-time pileup) are applied to this quantity on an event-by-event basis.

The invariant mass of each diphoton candidate (\mgg) is evaluated using the photon energies, the impact points measured in the calorimeter and the production vertex. The photon energy calibration is performed independently for converted and unconverted photons. Converted photons are defined to be those with a well-reconstructed conversion vertex in the inner detector. A detailed simulation of the detector geometry and response is used for the calibration. Additional corrections due to mis-modelling of the material in front of the calorimeter and of calorimeter non-uniformities are applied. These amount to about $\pm 1$\% depending on the pseudorapidity of the photon and are obtained from studies of \Zee\ decays in data~\cite{ElecPerfPaper2010}. The diphoton production vertex along the beam axis is determined by combining the trajectories of each photon, measured using the longitudinal segmentation of the calorimeter, with a constraint from the average beam spot position. The position of the conversion vertex is also used where the photons convert in the tracking region instrumented with silicon detectors. Conversion candidates with tracks reconstructed in inactive regions of the innermost pixel layer are rejected to reduce the contamination from misidentified electrons. The resolution of the diphoton mass reconstructed using this method is dominated by the photon energy resolution.

A total of 22,489 events were selected with a diphoton invariant mass between 100~GeV and 160~GeV. Although not used directly in the final result, the diphoton sample composition was studied using a two-dimensional side-band technique based on photon identification quality and isolation~\cite{Hgg_paper}. The fraction of true diphoton events was estimated to be $(71\pm 5)\%$. The rest of the background is due to events with one or more misidentified jets, except for a small ($\sim 0.7$\%) contribution from Drell-Yan events where both electrons pass the photon selection.

To enhance the sensitivity of the analysis, the data sample is split into nine categories, each with different expected signal mass resolutions, signal yields and signal-to-background ratios (S/B). This categorisation
depends on the impact point of the photons on the calorimeter, the presence of photon conversions and the value of the component of the diphoton transverse momentum orthogonal to the diphoton thrust-like axis in the transverse plane\footnote{$p_{\rm{Tt}} = {|\vec{p}_{\rm{T}}^{\gamma\gamma} \times \widehat{t}|}$, where ${\widehat{t} = \frac{ {\vec{p}_\mathrm{T}^{\gamma_1}} -{\vec{p}_\mathrm{T}^{\gamma_2}} }{ |{\vec{p}_\mathrm{T}^{\gamma_1}} - {\vec{p}_\mathrm{T}^{\gamma_2}}|}}$ denotes the transverse thrust, ${\vec{p}_\mathrm{T}^{\gamma_1}}$ and ${\vec{p}_\mathrm{T}^{\gamma_2}}$ are the transverse momenta of the two photons, and ${\vec{p}_{\rm{T}}^{\gamma\gamma}={\vec{p}_\mathrm{T}^{\gamma_1}} + {\vec{p}_\mathrm{T}^{\gamma_2}}}$ is the transverse momentum of the diphoton system.} (\ptt)~\cite{Ackerstaff:1997rc,Vesterinen:2008hx}.

Events in which both photons are unconverted are separated into the \emph{unconverted central} (both photons in the central region of the barrel calorimeter, $|\eta| < 0.75$) and \emph{unconverted rest} (all other events)
categories. Events for which at least one photon is converted are separated into the \emph{converted central} (both photons within $|\eta| < 0.75$), \emph{converted transition} (at least one photon close to the barrel/end-cap transition region, $1.3 < |\eta| < 1.75$)  and \emph{converted rest} (the remaining events) categories.

With the exception of the \emph{converted transition} category, all the events are further subdivided into \emph{low \ptt} ($\ptt < 40$~GeV) and \emph{high \ptt} (all other events) categories. Monte Carlo (MC) simulation studies show that a fermiophobic Higgs boson signal has larger \ptt\ on average than background events. This quantity is strongly correlated with the diphoton transverse momentum but offers several advantages. Higher values of \ptt\ do not include kinematic configurations for which the two photons are back-to-back in the azimuthal plane with substantially different transverse momenta. This reduces biases on the identification and isolation (transverse energy deposited around the photon) of the sub-leading photon in the high \ptt\ categories and retains a monotonically falling diphoton invariant mass distribution for the background events at the chosen cut values. The latter quality is advantageous for the background modelling and associated uncertainties discussed below.

A full {\tt Geant4}-based~\cite{geant4} MC simulation~\cite{atlasSim} of Higgs boson events decaying into two photons is used to model the expected signal. The signal yields are normalised to next-to-next-to-leading-order production cross sections~\cite{Ciccolini:2007jr,Ciccolini:2007ec,Bolzoni:2010xr,Ciccolini:2003jy,Brein:2003wg,Dittmaier:2011ti} and the branching ratios for the fermiophobic Higgs boson are calculated using {\tt HDECAY}~\cite{Djouadi:1997yw}. Higgs boson VBF production is simulated using {\tt POWHEG}~\cite{Nason:2009ai} interfaced with {\tt PYTHIA}~\cite{Sjostrand:2006za} for showering and hadronisation, while {\tt PYTHIA} is chosen for the VH processes. Pileup effects are simulated by overlaying each MC event with a variable number of simulated inelastic $pp$ collisions, taking into account the LHC bunch-train structure~\cite{MC11Tune}.

A set of corrections is applied to the simulated events in order to match the data-taking conditions. The simulated events are re-weighted to reproduce the distribution of the average number of interactions per bunch crossing reconstructed in the data, which has a mean value of about nine for the data sample used in this analysis. The energies of the simulated photons are smeared to account for differences observed in studies of the calorimeter resolution with \Zee\ decays. Calorimeter shower shapes used in the photon identification are slightly shifted to improve the agreement with the distributions observed with inclusive photons from data.

The number of fermiophobic Higgs bosons expected after candidate selection and the overall signal selection efficiency
for various values of \mH\ are shown in Table~\ref{tab:signal}.  The signal selection efficiency increases from 32\% to 42\% as the Higgs boson mass
increases from 110~GeV to 150~GeV.

\begin{table}[h!]
  \caption{\label{tab:yields} Expected signal mass resolution ($\sigma_\mathrm{CB}$ and FWHM in GeV, see text) and total number of signal events ($N_\mathrm{S}$) for $\mH = 120$~GeV for each of the nine analysis categories and for the inclusive case. Also shown for each category are the number of observed events ($N_\mathrm{D}$) in the diphoton mass range from 100~GeV to 160~GeV, and the expected signal-to-background ratio (S/B) in a mass window containing 90\% of the signal.}
  \small
  \myfontsizetwo
  \begin{center}
    \begin{tabular}{l@{}c@{\phantom{..}}c@{\phantom{..}}c@{\phantom{..}}c@{\phantom{..}}r@{\phantom{..}}r}
      \hline\hline
      \T \B Category & $\sigma_\mathrm{CB}$ & FWHM & $N_\mathrm{S}$ & $N_\mathrm{D}$ & S/B \\
      \hline
      \T \B Unconverted central, low \ptt   &  1.4  & 3.3 & 6.2 & 1763 & 0.03 \\
      \T \B Unconverted central, high \ptt  &  1.3  & 3.2 & 8.6 & 235 & 0.37  \\
      \T \B Unconverted rest, low \ptt      &  1.7  & 3.9 & 12.1 & 6234 & 0.02 \\
      \T \B Unconverted rest, high \ptt     &  1.6  & 3.8 & 16.0 & 1006 & 0.13  \\
      \T \B Converted central, low \ptt     &  1.6  & 3.8 & 4.0 & 1318 & 0.02 \\
      \T \B Converted central, high \ptt    &  1.5  & 3.5 & 5.8 & 184 & 0.26  \\
      \T \B Converted rest, low \ptt        &  2.0  & 4.6 & 11.8 & 7311 & 0.01 \\
      \T \B Converted rest, high \ptt       &  1.9  & 4.4 & 16.1 & 1072 & 0.09 \\
      \T \B Converted transition            &  2.3  & 5.8 & 10.8 & 3366 & 0.01 \\
      \hline
      \T \B All categories                  &  1.7  & 3.9 & 91.2 & 22489 & 0.03 \\
      \hline\hline
    \end{tabular}
  \end{center}
\end{table}

The signal is modelled as the sum of a core component, described by a \emph{Crystal Ball} (CB) function~\cite{CSB}, and a wider Gaussian component incorporating outlying events. The latter component typically accounts for less than 5\% of the signal. Table~\ref{tab:yields} lists the expected full-width-at-half-maximum (FWHM) and Gaussian width of the core component ($\sigma_\mathrm{CB}$) for each of the nine event categories. The expected number of signal events for $\mH = 120$~GeV, the number of background events in the diphoton mass range of 100~GeV to 160~GeV, and the signal-to-background ratio in a mass window containing 90\% of the signal are also shown. The main sensitivity to the fermiophobic production modes comes from the high \emph{\ptt} categories due to their enhanced signal yields and signal-to-background ratios. Figure~\ref{fig:fit_mH120} shows the signal diphoton mass distribution summed over the high \ptt\ categories for a Higgs boson mass of 120~GeV.

\begin{figure}[h!]
  \centering
  \includegraphics[width=0.45\textwidth]{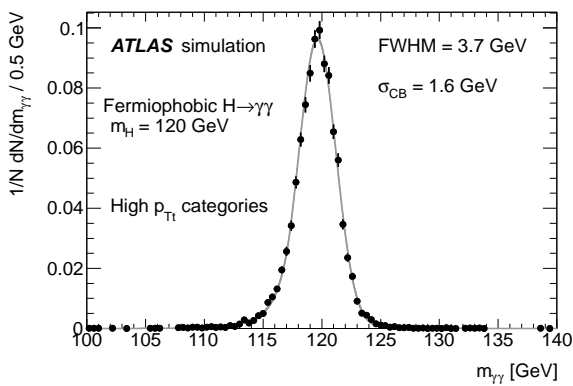}
  \caption{\label{fig:fit_mH120} Diphoton invariant mass spectrum from simulated signal samples (dots) with $\mH = 120$~GeV summed over the high \ptt\ categories, superimposed with the signal model (line).}
\end{figure}

The observed diphoton invariant mass distribution in each category is modelled by an exponential function. A fit to the data is performed for which the slope and normalisation are unconstrained.
Studies with large samples of simulated diphoton events show that this simple function gives a good description of the expected shape. The small systematic uncertainties associated with this assumption are discussed below.
Figures~\ref{fig:data_fit_cats}(a) and \ref{fig:data_fit_cats}(b) show the diphoton mass distributions of the selected data events summed over the low and high \ptt\ categories, respectively. The \emph{converted transition} category is included in the low \ptt\ categories.

\begin{figure}[h!]
 \includegraphics[width=0.45\textwidth]{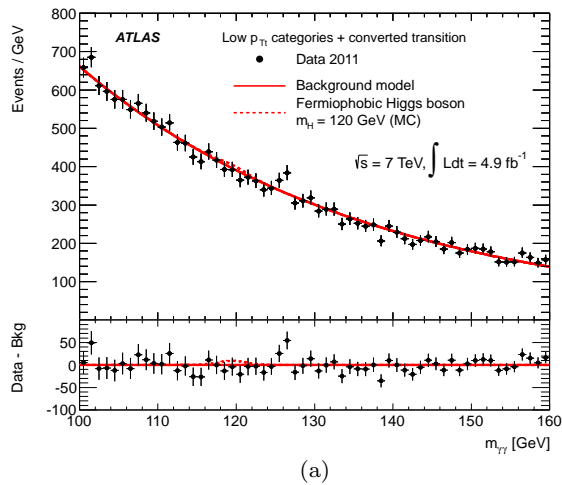}\vspace{-.5cm}
 \subfloat[][]{\hspace{0.50\textwidth}}\\ \vspace{.3cm}\\
 \includegraphics[width=0.45\textwidth]{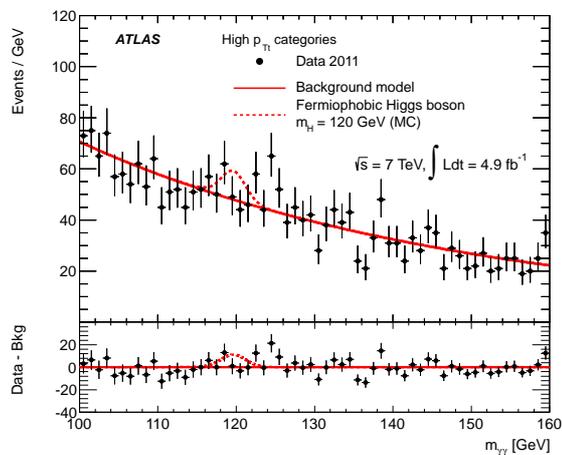}\\
  \vspace{-0.8cm} \\ \subfloat[][]{\hspace{0.50\textwidth}}
  \caption{\label{fig:data_fit_cats}Diphoton invariant mass spectra for the (a) low and (b) high \ptt\ categories, overlaid with the sum of the background-only fits from the individual categories. The bottom plots show the residual of the data with respect to the fitted background. The signal expectation for a fermiophobic Higgs boson with a mass of 120 GeV superimposed on the background fit is also shown.}
\end{figure}

Systematic uncertainties affecting the signal significance arise from uncertainties on the predicted signal yields, the expected partition of the signal among the categories and the modelling of the signal and background shapes. The dominant experimental uncertainty on the signal yield is due to the imperfect knowledge of the photon reconstruction and identification efficiencies, which is estimated to be $\pm 11$\%. This uncertainty is studied with electrons from $W$ and $Z$ boson decays in data, and photons from radiative decays of $Z$ bosons to electron and muon pairs.
In addition, the effect of pileup on photon identification gives a further contribution to the signal yield uncertainty of $\pm 4$\%. Uncertainties related to the trigger efficiency ($\pm 1$\%), isolation cut efficiency ($\pm 5$\%) and luminosity ($\pm 3.9$\%) are also included here.

Uncertainties on the signal cross section include a combination of the uncertainties on the parton distribution functions~\cite{Botje:2011sn,Alekhin:2011sk} and $\alpha_s$, and uncertainties on the QCD scale. Combining the VBF and VH production modes this uncertainty is within $\pm$4\% over the considered mass range.
To this uncertainty, that due to the \hgg\ branching ratio ($\pm$5\%) is added linearly, based on the SM calculation~\cite{Dittmaier:2011ti}.
This yields uncertainties of $\pm$9\% on the theoretical signal yield, leading to an overall uncertainty of $\pm 16$\% on the total signal expectation. In addition, the uncertainty on the Higgs boson \pt\ modelling is estimated by comparing signal samples from alternative
MC generators -- {\tt HERWIG}~\cite{Corcella:2002jc} for VBF and {\tt ResBos}~\cite{Balazs:1997xd}
for VH. The result is a $\pm$1\% signal migration between the low and high \ptt\ categories with a negligible effect on the signal selection efficiency.

The dominant uncertainties on the signal mass resolution are due to the uncertainty on the calorimeter energy resolution ($\pm$12\%) and photon calibration ($\pm$6\%), which are both extrapolated from the uncertainty on the electron calibration determined using $Z$ and $J/\psi$ data~\cite{ElecPerfPaper2010}.
The latter comes from the imperfect knowledge of the material in front of the active part of the calorimeter and is estimated using simulations with different amounts of material. This quantity also affects the fraction of expected events in the categories with converted photons; the maximal migration between \emph{converted} and \emph{unconverted} categories is estimated to be $\pm 4.5$\%. Other effects on the signal mass resolution are due to pileup fluctuations contributing to the cluster energy measurement ($\pm$3\%) and to the uncertainty on the photon angular resolution ($\pm$1\%)
which is studied in \Zee\ decays using the track-based direction measurement.
The total relative uncertainty on the diphoton invariant mass resolution is thus $\pm$14\%.

Systematic uncertainties on the background modelling arise from a possible deviation of the background mass distribution from the assumed exponential shape. This uncertainty is evaluated as the number of events that could be mistakenly attributed to the signal. It
is estimated from the adequacy of the chosen background
model's description of the mass distribution predicted
by {\tt ResBos}~\cite{Balazs:2007hr}. The residuals of the fit of the background
model to the {\tt ResBos} diphoton mass distribution are integrated
over a sliding mass window of 4 GeV, the approximate FWHM of the
expected signal.
The largest deviations were found at small invariant masses
and these uncertainties are then applied over the whole mass range. The resulting uncertainties range from $\pm 0.1$ to $\pm 7.9$ events in the individual analysis categories, where the magnitude of these uncertainties is roughly proportional to the number of background events in each category. These absolute uncertainties do not scale with the signal strength in the final likelihood fit. For a fermiophobic Higgs boson with $\mH = 120$~GeV the background modelling uncertainty in the high \ptt\ categories is equivalent to up to 5\% of the signal yield with nominal signal strength. The estimation of  the uncertainties is cross-checked by fitting the data with different functional forms and comparing the result to the exponential fit.

The possible presence of a signal is investigated using a combined likelihood function constructed from the signal and background models for the diphoton invariant mass distribution in each of the nine categories. Unbinned maximum likelihood fits of the signal strength are performed, treating the systematic uncertainties as nuisance parameters -- fourteen in total. These nuisance parameters are added to the signal likelihood function using a Gaussian term for the background modelling uncertainty, and log-normal terms for all other uncertainties.

\begin{figure}[!h]
	\centering
  \includegraphics[width=0.45\textwidth]{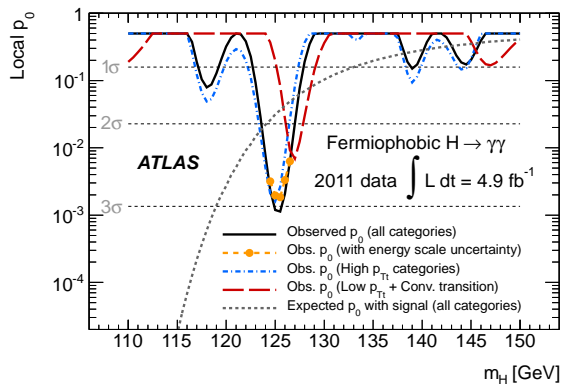}
	\caption{\label{fig:p0}Local observed $p_0$ as a function of the Higgs boson mass \mH\ (solid line) and the median expectation for a fermiophobic signal with the given \mH\ (dotted line).
The five points near 125~GeV show the observed $p_0$ when the uncertainty on the photon energy scale is considered. The individual contributions of the low \ptt\ and high \ptt\ categories to the observed $p_0$ are also shown.}
\end{figure}

The compatibility of the data with the background-only hypothesis, relative to the hypothesis of background plus the fermiophobic model signal, is quantified by the local significance $p_0$. Figure~\ref{fig:p0} shows the result for $m_H$ ranging from 110~GeV to 150~GeV, where $p_0$ is computed  in 0.5~GeV steps using asymptotic formulae~\cite{Cowan:2010js}.
The contributions to $p_0$ values from the high \ptt\ and low \ptt\ categories are shown separately.
The high \ptt\ contribution has a minimum $p_0$ at 125 GeV, while the low \ptt\ contribution
has a minimum at 127 GeV. The larger signal-to-background ratio as well as the larger expected
signal yield in the high \ptt\ category compared to the low \ptt\ category results in the high \ptt\ contribution dominating
in the final result.  The combined $p_0$ has a minimum at 125.5 GeV corresponding to 3.0 standard deviations.
The figure also shows the $p_0$ value expected for a fermiophobic Higgs boson signal, as a function of Higgs boson mass.

To obtain the final result, the impact of the uncertainties on the photon energy scale is considered for Higgs boson masses in the region of the minimum $p_0$, as shown in Fig.~\ref{fig:p0}. The corresponding effect on the
measured $p_0$ value is estimated using pseudo-experiments, since asymptotic formulae were found not to yield accurate estimates
of the probability in this case. The position of the minimum $p_0$ is almost unchanged and the significance is lowered to 2.9 standard deviations. Taking the look-elsewhere effect~\cite{LEE} into account
in the range 110 -- 150~GeV,
the significance reduces to about 1.6 standard deviations, with $p_0 \approx 0.051$. This may be compared to the result of a search for the SM Higgs boson performed with the same dataset and candidate selection~\cite{Hgg_paper}, yielding a minimum $p_0$ at a mass of 126.5~GeV with a global significance of 1.5~standard deviations. No statistically significant preference for either the SM or fermiophobic Higgs boson is observed.

Given the lack of evidence for a signal, mass-dependent exclusion limits on the fermiophobic benchmark model are calculated at the 95\% confidence level (CL) with a profile likelihood ratio test statistic in the $CL_s$ modified frequentist approach~\cite{Read:2002hq,HCG,Cowan:2010js} and are shown in Fig.~\ref{fig:CLsNorm}. Fermiophobic Higgs boson masses from 110.0~GeV to 118.0~GeV and from 119.5~GeV to 121.0~GeV are excluded, while the expected exclusion mass range is 110.0 -- 123.5~GeV. These results give more stringent lower mass limits than the previous results from LEP \\(108.2~GeV)~\cite{lep:2001yc} and the Tevatron (112.9~GeV from D0, 114~GeV from CDF)~\cite{Aaltonen:2009ga,Abazov:2011ix} in the diphoton decay channel.

\begin{figure}[!h]
	\centering
  \includegraphics[width=0.45\textwidth]{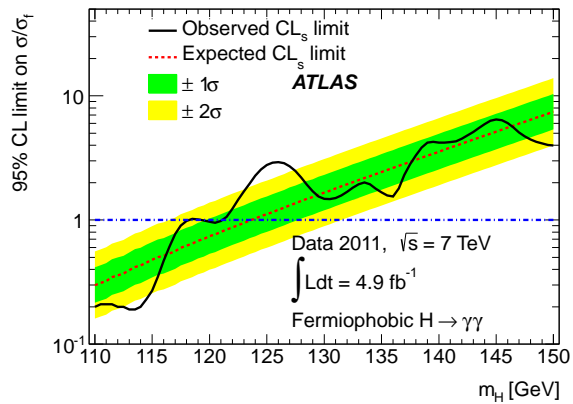}
	\caption{\label{fig:CLsNorm}Observed (solid line) and expected (dotted line) 95\% CL exclusion limits for a fermiophobic Higgs boson normalised to the fermiophobic cross section times branching ratio expectation ($\sigma_\mathrm{f}$) as a function of the Higgs boson mass hypothesis (\mH).}
\end{figure}

We thank CERN for the very successful operation of the LHC, as well as the
support staff from our institutions without whom ATLAS could not be
operated efficiently.

We acknowledge the support of ANPCyT, Argentina; YerPhI, Armenia; ARC,
Australia; BMWF, Austria; ANAS, Azerbaijan; SSTC, Belarus; CNPq and FAPESP,
Brazil; NSERC, NRC and CFI, Canada; CERN; CONICYT, Chile; CAS, MOST and NSFC,
China; COLCIENCIAS, Colombia; MSMT CR, MPO CR and VSC CR, Czech Republic;
DNRF, DNSRC and Lundbeck Foundation, Denmark; EPLANET and ERC, European Union;
IN2P3-CNRS, CEA-DSM/IRFU, France; GNAS, Georgia; BMBF, DFG, HGF, MPG and AvH
Foundation, Germany; GSRT, Greece; ISF, MINERVA, GIF, DIP and Benoziyo Center,
Israel; INFN, Italy; MEXT and JSPS, Japan; CNRST, Morocco; FOM and NWO,
Netherlands; RCN, Norway; MNiSW, Poland; GRICES and FCT, Portugal; MERYS
(MECTS), Romania; MES of Russia and ROSATOM, Russian Federation; JINR; MSTD,
Serbia; MSSR, Slovakia; ARRS and MVZT, Slovenia; DST/NRF, South Africa;
MICINN, Spain; SRC and Wallenberg Foundation, Sweden; SER, SNSF and Cantons of
Bern and Geneva, Switzerland; NSC, Taiwan; TAEK, Turkey; STFC, the Royal
Society and Leverhulme Trust, United Kingdom; DOE and NSF, United States of
America.

The crucial computing support from all WLCG partners is acknowledged
gratefully, in particular from CERN and the ATLAS Tier-1 facilities at
TRIUMF (Canada), NDGF (Denmark, Norway, Sweden), CC-IN2P3 (France),
KIT/GridKA (Germany), INFN-CNAF (Italy), NL-T1 (Netherlands), PIC (Spain),
ASGC (Taiwan), RAL (UK) and BNL (USA) and in the Tier-2 facilities
worldwide.

\bibliographystyle{spphys}       
\bibliography{FermiPaper}

\clearpage
\begin{flushleft}
{\Large The ATLAS Collaboration}

\bigskip

G.~Aad$^{\rm 48}$,
B.~Abbott$^{\rm 111}$,
J.~Abdallah$^{\rm 11}$,
S.~Abdel~Khalek$^{\rm 115}$,
A.A.~Abdelalim$^{\rm 49}$,
O.~Abdinov$^{\rm 10}$,
B.~Abi$^{\rm 112}$,
M.~Abolins$^{\rm 88}$,
O.S.~AbouZeid$^{\rm 158}$,
H.~Abramowicz$^{\rm 153}$
% ...
\bigskip

% ...
$^{10}$ Institute of Physics, Azerbaijan Academy of Sciences, Baku, Azerbaijan\\
$^{11}$ Institut de F\'isica d'Altes Energies and Departament de F\'isica de la Universitat Aut\`onoma  de Barcelona and ICREA, Barcelona, Spain\\
% ...
$^{48}$ Fakult\"{a}t f\"{u}r Mathematik und Physik, Albert-Ludwigs-Universit\"{a}t, Freiburg i.Br., Germany\\
$^{49}$ Section de Physique, Universit\'e de Gen\`eve, Geneva, Switzerland\\
% ...
$^{88}$ Department of Physics and Astronomy, Michigan State University, East Lansing MI, United States of America\\
% ...
$^{111}$ Homer L. Dodge Department of Physics and Astronomy, University of Oklahoma, Norman OK, United States of America\\
$^{112}$ Department of Physics, Oklahoma State University, Stillwater OK, United States of America\\
% ...
$^{115}$ LAL, Universit\'e Paris-Sud and CNRS/IN2P3, Orsay, France\\
% ...
$^{153}$ Raymond and Beverly Sackler School of Physics and Astronomy, Tel Aviv University, Tel Aviv, Israel\\
% ...
$^{158}$ Department of Physics, University of Toronto, Toronto ON, Canada\\
% ...
\end{flushleft}


\end{document}